\documentclass[11pt]{article}
\usepackage[dvips]{graphicx, color}
\usepackage{amsfonts}
\usepackage{amsmath}
\usepackage{amscd}
\usepackage{amssymb}
\usepackage{bm}
\usepackage[all]{xy}
\usepackage{enumerate}
\newcommand{\GSM}{G_{\textrm{SM}}}
\newcommand{\GpSM}{G'_{\textrm{SM}}}
\newcommand{\GtSM}{\tilde{G}_{\textrm{SM}}}
\newcommand{\itGamma}{\mathit{\Gamma}}
\newcommand{\Aut}{\textrm{Aut}}
\newcommand{\diag}{\textrm{diag}}

\newcommand{\Ker}{\textrm{Ker}}

\newcommand{\Id}{\textrm{Id}}
\newcommand{\Inn}{\textrm{Inn}}

\def\PRD#1#2#3{Phys.\ Rev.\ \textbf{D {#1}},  #2 (#3)}
\def\PRL#1#2#3{Phys.\ Rev.\ Lett.\ \textbf{{ #1}},  #2 (#3)}

\def\PL#1#2#3{Phys.\ Lett.\ \textbf{{#1}}, #2 (#3)}

\begin{document} 

\begin{flushright}

\end{flushright}

\vspace{3mm}

\begin{center}
{\Large \bf Semidirect product  gauge group 
$\left[SU(3)_{\rm c} \times SU(2)_{\rm L}\right]\rtimes U(1)_{\rm Y}$ 
\\
 and quantization of hypercharge }

\vspace{15mm}

Chuichiro Hattori $^{\rm{1},}$
            \footnote{E-mail: hattori@aitech.ac.jp}, 
Mamoru Matsunaga $^{\rm{2},}$
            \footnote{E-mail: matsuna@phen.mie-u.ac.jp}, and
Takeo Matsuoka $^{\rm{2},}$
            \footnote{E-mail: t-matsu@siren.ocn.ne.jp}

\end{center}

\begin{center}
\textit{
$^{\rm{1}}$Science Division, General Education, Aichi Institute of Technology, \\
     Toyota 470-0392, JAPAN \\
$^{\rm{2}}$Department of Physics Engineering, Mie University, \\
     Tsu 514-8507, JAPAN
 }
\end{center}

\vspace{5mm}

\begin{abstract}
In the Standard Model the hypercharges of quarks and leptons are not determined 
by the gauge group 
$SU(3)_{\rm c} \times SU(2)_{\rm L} \times U(1)_{\rm Y}$ alone. 
We show that, if we choose the semidirect product group 
$[SU(3)_{\rm c} \times SU(2)_{\rm L}] \rtimes U(1)_{\rm Y}$ as its gauge group, 
the hyperchages are settled to be $n/6 \mod {\mathbb{Z}}\;(n = 0,1,3,4) $. 
In addition, the conditions for gauge-anomaly cancellation give strong constraints. 
As a result, the ratios of the hypercharges are uniquely determined and 
the gravitational anomaly is automatically canceled.
The standard charge assignment to quarks and leptons can be properly 
reproduced. 
For exotic matter fields their hypercharges are also discussed. 
\end{abstract}

\newpage 
\section{INTRODUCTION}

The basic ingredients of the Standard Model are the gauge symmetry 
$\GSM = SU(3)_{\rm c} \times SU(2)_{\rm L} \times U(1)_{\rm Y}$ and 
the designation of its matter content. 
Quarks and leptons are assigned to $(\mathbf{3}, \ \mathbf{2}, \ 1/6)$, 
$(\mathbf{3}, \ \mathbf{1}, \ 2/3)$, $(\mathbf{3}, \ \mathbf{1}, \ -1/3)$, 
$(\mathbf{1}, \ \mathbf{2}, \ -1/2)$, and $(\mathbf{1}, \ \mathbf{1}, \ -1)$ 
representations of $\GSM$. 
These assignments are consistent with the cancellation of gauge anomalies 
\cite{anomaly}.
However, aside from the anomaly-free conditions, 
the direct product group $\GSM$ in itself does not give any relation 
between the $y$-charges ($U(1)_{\rm Y}$-charges) and 
the irreducible representations of $SU(3)_{\rm c} \times SU(2)_{\rm L}$. 
Even if we require the gauge-anomaly-free conditions on the $y$-charges under 
the above $SU(3)_{\rm c} \times SU(2)_{\rm L}$ assignments, 
the general solution of the $y$-charges contains two arbitrary real parameters. 
This means that there exist infinitely many solutions in which the 
$y$-charge assignments are quite different from those in the Standard Model. 
Within the context of the Standard Model with $\GSM$ we have no theoretical 
reason why  non-standard $y$-charge assignments are ruled out.

In order to single out the standard $y$-charge assignment, we have to rely 
on some kind of  selection rule. 
In view of the fact that, in general, a selection rule stems from a symmetry 
of the theory, it is appropriate that we modify the gauge group 
$\GSM$ in some manner. 
One possibility of the modification is to enlarge $\GSM$ to a simple group. 
The $SU(5)$ GUT is a typical example along this line of 
the modification \cite{SU(5)}.
As is well-known, through the traditional $SU(5)$ GUT-breaking in which 
an adjoint Higgs field acquires a non-zero vacuum expectation value along 
an appropriate direction, the standard charge assignment can be derived. 
However, the necessity of introducing the selection rule does not 
always require enlargement of the gauge group.

In this paper we consider another possibility of the modification in which 
the gauge group is non-simple and also its dimension remains to be 
${\rm dim} \, \GSM = 12$. 
Concretely, we choose the semidirect product gauge group 
$\GpSM = [SU(3)_{\rm c} \times SU(2)_{\rm L}] \rtimes U(1)_{\rm Y}$ and 
its linear representations. 
The semidirect product structure of the gauge group gives 
a selection rule:  the hypercharges are restricted to quantized values as 
$n/6\mod {\mathbb{Z}}\;(n = 0,1,3,4) $. 
Furthermore, the conditions for gauge-anomaly cancellation strongly 
constrain the $y$-charges of the matter fields and its solution contains 
only one arbitrary integer. 
As a result, the ratios of the $y$-charges are uniquely determined and 
the standard $y$-charge assignment could be properly reproduced.

Incidentally, if we incorporate an external gravitational field, we should take account of an additional 
constraint coming from the gravitational-anomay cancellation \cite{gravity1, gravity2}. 
In the Standard Model with $\GSM$, there are two types of solutions, in which one real parameter is left 
undetermined but the ratios of the $y$-charges are settled. 
One of the solutions accommodates the standard $y$-charge assignment but the other does not \cite{Weinberg}. (For original references, see \cite{Marshak, Minahan}.)
On the other hand, in the theory with $\GpSM$ the ratios of the $y$-charges are uniquely derived only from the gauge-anomaly cancellation condition and the gravitational anomaly is automatically canceled.

This paper is organized as follows. 
In Sec.~II we give a brief explanation of the semidirect product gauge group 
$\GpSM = [SU(3)_{\rm c} \times SU(2)_{\rm L}] \rtimes U(1)_{\rm Y}$. 
The linear representations of $\GpSM$ and the $y$-charge quantization 
are studied in Sec.~III. \;
Sec.~IV contains the gauge interactions and the solution 
of the anomaly-free conditions on the $y$-charges. 
In Sec.~V we mention alternative constructions of gauge groups.
In Sec.~VI we discuss possible charge assignments for exotic matter fields 
which are in line with the anomaly-free conditions. 
Sec.~VII concludes with a brief summary of our results.


\section{SEMIDIRECT PRODUCT GROUP 
$\GpSM =[SU(3)_{\rm c} \times SU(2)_{\rm L}] \rtimes U(1)_{\rm Y}$}

As stated in the introduction, we consider the semidirect product gauge group 
$\GpSM =[SU(3)_{\rm c} \times SU(2)_{\rm L}] \rtimes U(1)_{\rm Y}$.

In general, a semidirect product of two groups $G$ and $H$, $G \rtimes H$, 
is defined as a set of pairs $(g, h)$ ($g\in G,\, h \in H$) endowed 
with the product rule \cite{Ramond}
\begin{equation}
	  (g, h) (g',  h') = (g \, \sigma_h(g'), \, hh'). 
	\label{eq:SDP}
\end{equation}
Here $\sigma_h$ is an automorphism of the group $G$ for each $h \in H$ 
which satisfies 
\begin{equation}
	\sigma_{hh'} = \sigma_h \circ \sigma_{h'}\, . 
\end{equation}
This means that the map $\sigma : H  \ni h \mapsto   \sigma_h \in \Aut (G)$ 
is a homomorphism. 
We notice that the elements of $H$ participate in the product of the elements 
of $G$. 
There exist several semidirect products, each corresponding to a choice of 
this homomorphism $\sigma$. 
When $\sigma$ is trivial, i.e.,
$\sigma_h = \Id \; (\forall h \in H)$, $G \rtimes H$ is nothing but 
the direct product group $G \times H$, which contains both $G$ and $H$ 
as normal subgroups.

For $G = G_{32} :=SU(3) \times SU(2)$ and $H = U(1)$, it can be easily shown 
that a nontrivial homomorphism from $H$ to $\Aut (G_{32})$ is that from $H$ 
to the inner automorphism group $\Inn(G_{32})$, which is expressed as 
\begin{equation}
	\begin{split}
	   \imath : U(1) \ni e^{i\theta}  \mapsto \imath_{\theta}(\bullet) 
	      &= \gamma(\theta)\, (\bullet) \, \gamma(\theta) ^{-1} \in \Inn(G_{32}), \\
              & \gamma(\theta) = (\gamma_3(\theta),\gamma_2(\theta)) \in G_{32}=SU(3) \times SU(2). 
	\label{eq:Inner}
	\end{split}
\end{equation}
In order to get a semidirect product Lie group, we require the following conditions for $\imath$:
(i) continuity,
(ii) well-definedness \;$ \imath_{\theta+2\pi}= \imath_{\theta}$,
(iii)  homomorphism \;$\imath_{\theta}\,\imath_{\theta'} = \imath_{\theta +\theta'}$.

We rewrite the conditions (i)--(iii) in terms of $\gamma(\theta)$.
Notice that, if  $\gamma(\theta)$ and $\gamma'(\theta)$ differ up to the center $Z(G_{32})$, i.e.,
\begin{equation}
	\gamma(\theta) = \alpha(\theta)\gamma'(\theta),\quad \alpha(\theta) \in Z(G_{32}) ,
\end{equation}
then  $\gamma(\theta)$ and $\gamma'(\theta)$ define the same homomorphism $\imath$. 
As to the condition (i), it is obvious that we can always choose a continuous $\gamma(\theta)$ for each continuous  $\imath_{\theta}$. Hence we may think $\gamma(\theta)$ continuous.
The condition (ii) and Eq.~\eqref{eq:Inner} lead to 
\begin{equation}
    \gamma(\theta +2\pi) = c(\theta)\,\gamma(\theta), \quad 
       c \in  Z(G_{32}) = \mathbb{Z}_3 \times \mathbb{Z}_2.  \label{c} 
\end{equation}
Since  $\gamma(\theta)$ is a continuous map and $Z(G_{32})$ is a finite set, the factor $ c(\theta)$ is independent of $\theta$.
The contidion (iii) means 
\begin{equation}
  \gamma(\theta) \gamma(\theta') = f(\theta, \theta') \gamma(\theta + \theta'),  
             \quad   f(\theta ,\theta' ) \in Z(G_{32}) .
	\label{eq:product}
\end{equation}
The set $\{ f(\theta ,\theta')\}$ is called a factor set or a factor 
system \cite{Azcarraga}. 
Setting $\theta' =0$ in Eq.~\eqref{eq:product}, we get $\gamma(0)= f(\theta, 0)$. From the continuity of $\gamma(\theta)$ and the finiteness of $Z(G_{32})$, we find that $f(\theta ,\theta' )$ is independent of $\theta$ and $\theta'$. Hence,  $f(\theta ,\theta' ) = \gamma(0)$.
Furthermore, by the redefinition $\gamma(\theta) \rightarrow \gamma(0)\gamma(\theta)$, we can set $\gamma(0)=\mathbf{1}$.

We are thus led to the conditions
\begin{align}
    \gamma(\theta +2\pi) &= c\,\gamma(\theta), \quad 
       c = ( e^{i\frac{2\pi}{3}m}\mathbf{1}_3,  e^{-i\pi n}\mathbf{1}_2), \quad m=0,\pm1,\,n=0,1 , 
\label{eq:quasi-per}  \\
	 \gamma(\theta) \gamma(\theta') &= \gamma(\theta + \theta') .
\label{eq:homo}
\end{align}  
The general solution of Eqs.~\eqref{eq:quasi-per} and \eqref{eq:homo}, which we denote $\tilde{\gamma}(\theta) =\left(\tilde{\gamma}_3(\theta),\tilde{\gamma}_2(\theta)\right) \in SU(3) \times SU(2)$, is of the form
\begin{equation}
	\begin{split}
	    \tilde{\gamma}_3^{(m,m')}(\theta) 
	    &= U_3 \,e^{i\left(\frac{m}{3}+m'\right)\theta\Lambda_8}\,{U_3}^{\dag},\quad m=0,\pm1,\quad\Lambda_8
	    = \diag(1, 1, -2),
	 \\
   	 \tilde{\gamma}_2^{(n,n')}(\theta ) 
	 &= U_2\,e^{-i\left(\frac{n}{2}+n'\right)\theta\tau_3}\,{U_2}^{\dag},\quad n=0,1,\quad\tau_3 = \diag(1, -1),
	\end{split}
\end{equation}
where  $m'$ and $n'$ are arbitrary integers and $U_N\; (N=2,3)$ are arbitrary $SU(N)$-matrices. 
We note that  $U_N\; (N=2,3)$ are independent of $\theta$, because $\gamma(\theta)$ and $\gamma(\theta')$ are commutative due to Eq.~\eqref{eq:homo} and hence simultaneously diagonalizable by the same $SU(N)$-matrices  $U_N(\theta) = U_N(\theta') = U_N(0)$.

Each of the above solutions $\tilde{\gamma} = (\tilde{\gamma}_3^{(m,m')},\tilde{\gamma}_2^{(n,n')})$ defines the semidirect product $G_{32} \rtimes_{\tilde{\imath}}U(1)$ through the homomorphism $\tilde{\imath}_{\theta}(\bullet)  = \tilde{\gamma}(\theta)\, (\bullet) \, \tilde{\gamma}(\theta) ^{-1}$. There exist many semidirect products.
However, it can be shown that the semidirect product $G_{32} \rtimes_{\tilde{\imath}} U(1)$ defined through $(\tilde{\gamma}_3^{(m,m')},\tilde{\gamma}_2^{(n,n')})$ is isomorphic to the semidirect product $G_{32} \rtimes_{\imath} U(1)$ defined through $\gamma(\theta) = (\gamma^{(m)}_3(\theta),\gamma^{(n)}_2(\theta))$ with
\begin{equation}
	\begin{split}
 	  \gamma^{(m)}_3(\theta) &= e^{\frac{m}{3}i \theta \Lambda_8}, 
	       \quad m=0,\pm1, \\
	   \gamma^{(n)}_2(\theta) &= e^{-\frac{n}{2}i \theta \tau_3}, 
       \quad n=0,1.
	\end{split}
\end{equation}
Indeed, we can verify that the map
\begin{equation}
	F\; : \; G_{32} \rtimes_{\imath} U(1) \ni  
	 (g,e^{i\theta}) \mapsto  (\tilde{g},e^{i\tilde{\theta}}) \in G_{32} \rtimes_{\tilde{\imath}} U(1) 
\end{equation}
defined by
\begin{align*}
	\tilde{g} &= \phi(g, \theta) = g\gamma(\theta) {\tilde{\gamma}(\theta)}^{-1},\\
	\tilde{\theta} &=\theta
\end{align*}
gives the Lie-group isomorphism in the following way.
First, this map is well-defined, i.e.,\,$\phi(g, \theta +2\pi)=\phi(g, \theta)$. 
Second, it is obviously bijective and analytic.
Finally, it is a homomorphism because we have by the definition of $F$  
\begin{align}
	F\bigl((g,e^{i \theta})(g',e^{i \theta'})\bigr) 
	& =F\bigl(g \imath_{\theta} (g') , e^{i \theta}e^{i \theta'}\bigr) = \bigl(\phi\bigl(g \imath_{\theta}  (g'), \theta + 		\theta'\bigr),  e^{i (\theta + \theta')}\bigr), 
		\label{eq:F1}\\
	F(g,e^{i \theta}) F(g',e^{i \theta'}) &=\bigl(\phi (g, \theta ), e^{i \theta}\bigr) \bigl(\phi (g', \theta' ), e^{i \theta'}\bigr)
	=  \bigl( \phi(g,\theta)\tilde{\imath}_{\theta} \bigl(\phi (g',\theta')\bigr),  e^{i (\theta + \theta')} \bigr)
		\label{eq:F2}
\end{align}
and a straightforward calculation confirms the equality of the first components of Eqs.~\eqref{eq:F1} and \eqref{eq:F2}.

Furthermore, because of the map $\theta \mapsto -\theta$ gives the automorphism of the group $U(1)$, we find that $(\gamma^{(m =-1)}_3,\gamma^{(n)}_2)$ and $(\gamma^{(m =1)}_3,\gamma^{(n)}_2)$ define the semidirect product groups isomorphic to each other.

To sum up, there exist  four essentially different semidirect product groups defined through $(\gamma^{(0)}_3,\gamma^{(0)}_2)$, $(\gamma^{(0)}_3,\gamma^{(1)}_2)$, $(\gamma^{(1)}_3,\gamma^{(0)}_2)$, and $(\gamma^{(1)}_3,\gamma^{(1)}_2)$.  They are 
$SU(3) \times SU(2) \times U(1),\; 
SU(3) \times [ SU(2) \rtimes U(1)],\; 
SU(2) \times [ SU(3) \rtimes U(1)]$,
and $\left[SU(3) \times SU(2)\right]\rtimes U(1)$, respectively. 
Among them the last one, 
$\GpSM =[SU(3)_{\rm c} \times SU(2)_{\rm L}] \rtimes U(1)_{\rm Y}$, 
 could be able to constrain hypercharges of both quarks and leptons. 

In the following, we concentrate our attention to this semidirect produt and write simply $\left(\gamma_3,\gamma_2\right)$ instead of $(\gamma^{(1)}_3,\gamma^{(1)}_2)$.
The product rule of $\bigl(g_3, g_2, e^{i\theta}\bigr),\,\bigl(g'_3, g'_2, e^{i\theta'}\bigr) \in \GpSM$ is 
\begin{equation}
	\bigl(g_3, g_2, e^{i\theta}\bigr)\bigl(g'_3, g'_2, e^{i\theta'}\bigr) 
	= \bigl(g_3 \gamma_3(\theta) g'_3{\gamma_3(\theta)}^{-1}, \, 
	g_2 \gamma_2(\theta) g'_2{\gamma_2(\theta)}^{-1}, \,e^{i(\theta + \theta')}\bigr).
\end{equation}

\section{ $\GpSM$ AND THE VALUES OF HYPERCHARGE}

We first construct a linear representation $\mathcal{R}$ of  $\GpSM = G_{32} \rtimes U(1) \ni (g, e^{i\theta})$. The commonly used method is to start from a representation $R$ of its subgroup  $G_{32}$. We note that the map $\GpSM  \ni (g,e^{i\theta}) \mapsto R\left(g \gamma(\theta)\right) \in GL_N(\mathbb{C})$  forms a \textit{projective} representation of $\GpSM $.
Indeed, the product of the  images  of two elements $(g,e^{i\theta}),(g',e^{i\theta'}) \in \GpSM$ is 
\begin{equation}
	R\left(g \gamma(\theta) \right) R\left(g' \gamma(\theta') \right) 
	= R \left( g \gamma(\theta)\, g' \gamma(\theta') \right)
	=R \left( g\imath_{\theta} (g')\,\gamma(\theta)\gamma(\theta')\right)
	=R\left( g\imath_{\theta} (g')\,\gamma(\theta + \theta') \right).
\end{equation}
However, $R\left(g \gamma(\theta)\right)$ is not periodic in $\theta$:
\begin{equation}
	R\left(g \gamma(\theta+2\pi) \right) =R(c)\,R\left(g \gamma(\theta) \right) .
\end{equation}

From these facts, we immediately find that, if we take a projective representation $\rho$ of $U(1)$ satisfying $\rho(\theta + 2\pi) = c\,\rho(\theta)$, the tensor product
\begin{equation}
  \mathcal{R}\bigl(g, e^{i\theta}\bigr)  
      = R \left(g \,\gamma (\theta)\right) \otimes \rho (\theta )^{\ast}
	\label{eq:tensor}
\end{equation} 
forms a linear representation of $\GpSM $. The cancellation of two factors $R(c)$ and $c$, respectively arising from  $R\left(g \gamma(\theta)\right)$ and $\rho (\theta )$, assures that $\mathcal{R}$ is a linear representation. 

It can be also shown that any linear representation of $\GpSM $ takes the above form. The proof is essentially based on the observation  that  any element $(g, e^{i\theta}) \in \GpSM $ is decomposed as
\begin{equation}
	\bigl(g, e^{i\theta}\bigr)
	= \bigl(g\gamma(\theta),1\bigr)\, \bigl(\gamma(\theta)^{-1}, e^{i\theta}\bigr)
	=\bigl(\gamma(\theta)^{-1}, e^{i\theta}\bigr) \,\bigl(g\gamma(\theta),1\bigr).
		\label{eq:decomp0}
\end{equation}
The tensorial nature of the linear representation \eqref{eq:tensor} comes from the commutativity of two factors  $\bigl(g\gamma(\theta),1\bigr)$ and $\bigl(\gamma(\theta)^{-1}, e^{i\theta}\bigr)$.
Non-periodic nature of the latter factor 
\begin{equation}
	\bigl(\gamma(\theta + 2\pi )^{-1}, e^{i(\theta + 2\pi)}\bigr)
	 =\big(c^{-1}\gamma(\theta)^{-1}, e^{i\theta}\bigr)
\end{equation}
leads to a projective representation $\rho$.

An irreducible representation 
$R$ of $SU(3)_{\rm c} \times SU(2)_{\rm L} \ni (g_3, g_2)$ 
is described in terms of the tensor product $R =R_3\otimes R_2$, 
with $R_3 \left( R_2\right)$ being an irreducible representation of 
$SU(3)_{\rm c} \left( SU(2)_{\rm L} \right)$. 
An irreducible projective representation of $U(1)_{\rm Y}$ is one-dimensional 
and written as $\rho (\theta ) = e^{- i y \theta}$. 
Hence an irreducible representation $\mathcal{R}$ of 
$\GpSM= \left[SU(3)_{\rm c} \times SU(2)_{\rm L}\right]\rtimes U(1)_{\rm Y} \ni 
\left(g_3, g_2, e^{i\theta}\right) $ is of the form
\begin{equation}
   \mathcal{R}\bigl(g_3, g_2, e^{i\theta}\bigr)
       = R _3\left(g _3\,\gamma_3 (\theta)\right) 
         \otimes R _2\left(g _2\, \gamma_2 (\theta)\right) 
            \otimes e^{i y \theta}.
	\label{eq:rep321}
\end{equation}
The non-periodicity factor  $R(c)$ is found as 
\begin{equation}
\begin{split}
    R\left(\gamma(\theta + 2\pi)\right) 
      & = R_3\left(\gamma_3(\theta + 2\pi)\right) 
          \otimes R_2\left(\gamma_2(\theta + 2\pi)\right) \\
            & = (e^{\frac{2\pi}{3}i})^{r_3}R_3\left(\gamma_3(\theta )\right) 
               \otimes (e^{-\pi i})^{r_2}R_2\left(\gamma_2(\theta)\right)  \\ 
   & = e^{- 2\pi i (\frac{r_2}{2} - \frac{r_3}{3})}R\left(\gamma(\theta)\right),
\end{split}
\end{equation}
where 
\[
	\begin{split}
	r_2  &:=2 I \qquad \qquad  I: \textrm{isospin},  \\
	r_3  &:=\lambda_1 + 2 \lambda_2 \qquad 
               [ \lambda_1, \lambda_2 ]_D : \textrm{the Dynkin label pair}. 
	\end{split}
\]
Consequently, we obtain the hypercharge \textit{quantization} as
\begin{equation}
    y  \equiv \frac{r_2}{2} - \frac{r_3}{3} \mod {\mathbb{Z}},
\label{eq:hypercharge}
\end{equation}
whose values are shown for each matter multiplet in TABLE~\ref{TABLE I}. 
It should be noted that the values of $y$ are determined up to additive 
integers.

\begin{table}[htdp]
\caption{Hypercharges $y$ in $\GpSM$.  $\Phi$ represents the Higgs field.}
\begin{center}\begin{tabular}{c  c c c c c c } \hline\hline
                & $ q_\textrm{L}$ & $u_\textrm{R}$ & $d_\textrm{R}$ 
                    & $ l_\textrm{L} $ &  $e_\textrm{R}$ & $\Phi $  \\
                & $\quad\mathbf{ (3, 2)}\quad$ & $\quad\mathbf{ (3, 1)}\quad$  & $\quad\mathbf{ (3, 1)}\quad$ 
                    & $\quad\mathbf{ (1, 2)}\quad$ & $\quad\mathbf{ (1, 1)}\quad$ &$\quad\mathbf{ (1, 2)}\quad$ \\
\hline
 $y \mod {\mathbb{Z}}$ & $\frac{1}{6}$ &  $-\frac{1}{3}$  &  $ -\frac{1}{3}$  
                      & $ \frac{1}{2}$  & $ 0 $& $ \frac{1}{2}$ \\
\hline\hline
\end{tabular} 
\end{center}
\label{TABLE I}
\end{table}

\section{GAUGE INVARIANCE AND ANOMALIES}

Let us consider a fermion field $\Psi$ which belongs to the representation $\mathcal{R}$ 
in $\GpSM$, Eq.~\eqref{eq:rep321}. 
The Lagrangian for $\Psi$ is of the form 
\[
    \mathcal{L}_{\Psi} = \bar{\Psi}\,i\,\gamma^{\mu} 
           \left(\partial_{\mu} - i \mathcal{A}_{\mu}\right)\, \Psi, 
\]
where the gauge field $\mathcal{A}_{\mu}$ is given by 
\begin{equation}
  \mathcal{A}_{\mu} = A^{(3)}_{\mu}  \otimes \mathbf{1}_2 
                       + \mathbf{1}_3 \otimes A^{(2)}_{\mu}  
                        + y \mathbf{1}_3 \otimes \mathbf{1}_2  A^{(1)}_{\mu}. 
\end{equation}
In order for $\mathcal{L}_{\Psi}$ to be invariant under the gauge transformation 
\[
    \Psi \rightarrow \Psi' = \mathcal{R} \Psi, 
\]
we put the transformation rule as 
\begin{equation}
\left\{
\begin{split}
   {A^{(3)}}'_{\mu} & = R_3\, A^{(3)}_{\mu}\,R_3^{\dag} 
                       - i \left(\partial_{\mu} R_3 \right)R_3^{\dag}, \\
   {A^{(2)}}'_{\mu} & = R_2\, A^{(2)}_{\mu}\,R_2^{\dag} 
                       - i \left(\partial_{\mu} R_2 \right)R_2^{\dag}, \\
   {A^{(1)}}'_{\mu} & = A^{(1)}_{\mu} + \partial_{\mu}\theta, 
\end{split}
\right.
\end{equation}
where $R_3 = R_3\left(g_3\,\gamma_3(\theta )\right)$ and 
$R_2 = R_2\left(g_2\,\gamma_2(\theta )\right)$. 
It should be noted that the gauge fields $A^{(3)}_{\mu}$ and $A^{(2)}_{\mu}$
 are affected also by the $U(1)_{\rm Y}$ transformation. 
This situation arises from the fact that the $U(1)_{\rm Y}$ transformation 
is at work on $SU(3)_{\rm c}$ and $SU(2)_{\rm L}$ 
through $\gamma_3 (\theta ) \in SU(3)_{\rm c} $ and 
$\gamma_2 (\theta ) \in SU(2)_{\rm L}$. 
The requirement of the invariance for any element 
$(g_3, g_2, e^{i\theta}) \in \GpSM$ is equivalent to that for 
any elements given by $(g_3, \mathbf{1}_2, 1)$, $(\mathbf{1}_3, g_2, 1)$, and 
$(\mathbf{1}_3, \mathbf{1}_2, e^{i\theta})$.  
Consequently, the Lagrangian in $\GpSM$ is the same as in $\GSM$.

As seen in TABLE~\ref{TABLE I}, the $y$-charges of the matter fields are fixed  
up to additive integers. 
The number of arbitrary additive integers is reduced 
by requiring the anomaly-free conditions. 
To see this, we proceed to study these conditions. 
Let us denote $y$-charges of the matter fields $q_{\rm L}$, 
$u_{\rm R}$, $d_{\rm R}$, $l_{\rm L}$, and $e_{\rm R}$ 
by $y(q_{\rm L})$, $y(u_{\rm R})$, $y(d_{\rm R})$, $y(l_{\rm L})$, 
and $y(e_{\rm R})$, respectively. 
The mixed anomaly conditions $U(1) \cdot (SU(3))^2$ and 
$U(1) \cdot (SU(2))^2$ are expressed as 
\begin{align}
	  & 2 y(q_{\rm L}) - y(u_{\rm R}) - y(d_{\rm R}) = 0, 
	\label{eq:mixed3}  \\
	  &                3 y(q_{\rm L}) + y(l_{\rm L}) = 0, 
	\label{eq:mixed2}
\end{align}
respectively. 
For the cubic $U(1)$ anomaly we obtain the constraint 
\begin{equation}
	  6 y(q_{\rm L})^3 - 3 y(u_{\rm R})^3 - 3 y(d_{\rm R})^3 
	    + 2 y(l_{\rm L})^3 - y(e_{\rm R})^3 = 0. 
\label{eq:cubic}
\end{equation}
When we use the notation $y(u_{\rm R}) = x$ and $y(d_{\rm R}) = z$, 
Eqs (\ref{eq:mixed3}) and (\ref{eq:mixed2}) are written as 
\begin{equation}
	y(q_{\rm L}) =   \frac{1}{2} ( x + z ), \qquad 
	y(l_{\rm L}) = - \frac{3}{2} ( x + z ). 
\label{eq:yql}
\end{equation}
Inserting Eq.~\eqref{eq:yql} into Eq.~\eqref{eq:cubic}, 
we can rewrite Eq.~\eqref{eq:cubic} as 
\begin{equation}
 	 6 (x + z)^3 + 3 x^3 + 3 z^3 + y(e_{\rm R})^3 = 0. 
	\label{eq:xz3}
\end{equation}
Through the linear transformation 
\[
	x = \frac{1}{3} ( -X + 2Z ), \qquad z = \frac{1}{3} ( 2X - Z ), 
\]
Eq.~\eqref{eq:xz3} is put into the  form 
\begin{equation}
	X^3 + Z^3 + y(e_{\rm R})^3 = 0. 
	\label{eq:Fermat}
\end{equation}
In the previous section it is found that $X$, $Z$, and 
$y(e_{\rm R})$ are rational numbers. 
Hence, Fermat's theorem asserts that 
if $X Z y(e_{\rm R}) \neq 0$, 
there is no rational solution of Eq.~\eqref{eq:Fermat}. 
This means 
\begin{equation}
	X \, Z \, y(e_{\rm R}) = 0. 
	\label{eq:Zero}
\end{equation}
The next step of our study is to find the solution of 
Eq.~\eqref{eq:Zero}.

As discussed in the previous section, $y(q_{\rm L})$ is of the form 
\[
	y(q_{\rm L}) = \frac{1}{6} + n, \qquad n \in \mathbb{Z}.
\]
It follows that $x + z = \frac{1}{3} + 2n$. 
Furthermore, $y$-charges of $u_{\rm R}$ and $d_{\rm R}$ are settled as 
$x \equiv z \equiv - \frac{1}{3} \mod{\mathbb{Z}}$. 
Consequently, $x$ and $z$ are given by 
\[
	x = \frac{2}{3} + n + s, \qquad z = -\frac{1}{3} + n - s 
\]
with $s \in \mathbb{Z}$. 
Thus we find 
\[
	X = x + 2z = 3n - s, \qquad Z = 2x + z = 3n + s + 1. 
\]
Let us remember that, as shown in Eq.~\eqref{eq:Zero}, one of $X$, $Z$, and $y(e_{\rm R})$ should be zero. 
If $y(e_{\rm R}) = 0$, then, from Eq.~\eqref{eq:Fermat},  we have 
$X + Z = 0$. 
This leads to the relation $6n + 1 = 0$ 
inconsistent with $n \in \mathbb{Z}$. 
Therefore, the solution of Eq.~\eqref{eq:Zero} becomes 
$X = 0$ or $Z = 0$. 
In the case of $X = 0$ we obtain $s = 3n$ and $y(e_{\rm R}) = - Z$. 
The result is 
\begin{align}
	y(q_{\rm L}) & = \frac{1}{6} + n, \nonumber \\
	y(u_{\rm R}) & =  4 y(q_{\rm L}) = \frac{2}{3} + 4n, \nonumber \\
	y(d_{\rm R}) & = -2 y(q_{\rm L}) = -\frac{1}{3} - 2n, \\
	y(l_{\rm L}) & = -3 y(q_{\rm L}) = -\frac{1}{2} - 3n, \nonumber \\
	y(e_{\rm R}) & = -6 y(q_{\rm L}) = -1 - 6n, \nonumber 
\end{align}
with $n \in \mathbb{Z}$. 
In the case of $Z = 0$ we obtain $s = - 3n - 1$ and $y(e_{\rm R}) = - X$. 
Aside from  the interchange of $u_{\rm R}$ and $d_{\rm R}$, 
the result is the same as in the case of $X = 0$. 
In TABLE~\ref{TABLE II} the $y$-charges obtained here are shown. 
Although the integer $n$ is undetermined, 
it is remarkable that the ratios of the $y$-charges are completely fixed.

\begin{table}[htdp]
\caption{Hypercharges $y$ from the anomaly-free conditions in $\GpSM$ 
$(n \in \mathbb{Z})$ }
\begin{center}\begin{tabular}{c c c c c c c} \hline\hline
	&$ q_\textrm{L}$ & $u_\textrm{R}$ & $d_\textrm{R}$ 
		&$ l_\textrm{L} $ &  $e_\textrm{R}$ & $\Phi $  \\
	& $\quad\mathbf{ (3, 2)}\quad$ & $\quad\mathbf{ (3, 1)}\quad$  & $\quad\mathbf{ (3, 1)}\quad$ 
		& $\quad\mathbf{ (1, 2)}\quad$ & $\quad\mathbf{ (1, 1)}\quad$ &$\;\mathbf{ (1, 2)}\quad$ \\
\hline
	$y$ & $\frac{1}{6} + n $ &  $ \frac{2}{3} + 4n $  &  $ -\frac{1}{3} -2n $  
		 & $ - \frac{1}{2} -3n $  & $ -1- 6n $& $ \frac{1}{2}$ \\
\hline\hline
\end{tabular} 
\end{center}
\label{TABLE II}
\end{table}

The $y$-charge of the Higgs field $\Phi$ which belongs to $\mathbf{ (1, 2)}$ 
representation is given by $y_{\Phi} = \frac{1}{2} + m$ $(m \in \mathbb{Z})$. 
We assume the spontaneous symmetry breaking 
$\left[SU(3)_{\rm c} \times SU(2)_{\rm L}\right]\rtimes U(1)_{\rm Y} 
\longrightarrow SU(3)_{\rm c} \rtimes U(1)_{\rm em}$ through 
a non-zero value of $\langle \Phi \rangle$. 
The NNG-relation 
$Q_{\rm em} = I_3 + Y$  leads to $y_{\Phi} = 1/2$, i.e., $m = 0$. 
Additionally, if we suppose $Q_{\rm em}(e_{\rm R}) = - 1$, we obtain $n = 0$. 
As a result, $y$-charges of the matter fields are completely determined. 
The solution is nothing but the Standard Model charge assignment. 
Thus we have Yukawa couplings for quarks and leptons.

As mentioned above, in the semidirect product gauge group $\GpSM$ 
the ratios of $y$-charges are uniquely determined under the conditions for 
the gauge-anomaly cancellation. 
Furthermore, in the solution the gravitational anomaly is \textit{automatically 
canceled}. Indeed  the mixed $U(1)({\rm graviton})^2$ anomaly is canceled as seen from the TABLE~\ref{TABLE II}:
\begin{equation}
	6 \, y(q_{\rm L}) - 3 \, y(u_{\rm R}) - 3 \, y(d_{\rm R}) 
		 + 2 \, y(l_{\rm L}) - y(e_{\rm R}) = 0. 
\label{eq:gravity}
\end{equation}

This is in marked contrast to the Standard Model with $\GSM$,
in which the ratios of $y$-charges are not determined under the gauge-anomaly condition and may vary continuously since $y$-charges are not restricted to rational numbers.
We are to take account of the cancellation of gravitational anomaly, i.e., Eq.~\eqref{eq:gravity}, as an \textit{additional constraint} on the $y$-charges \cite{Weinberg}. Translating Eq.~\eqref{eq:gravity} into  the form
\begin{equation}
	X + Z + y(e_{\rm R}) = 0
\end{equation}
and combining this with Eq.~\eqref{eq:Fermat}, we obtain 
\[
	X Z y(e_{\rm R}) = 0, 
\]
which is just the same as Eq.~\eqref{eq:Zero} except that $X, Z$, and $y(e_{\rm R})$ are not necessarily rational numbers.
This equation allows two types of the solution in which the ratios of $y$-charges 
are settled. 
One of them accommodates the standard $y$-charge assignment but the other does not. 

\section{GAUGE GROUPS OTHER THAN $\GpSM$}

The semidirect product group $\GpSM$ constructed in Sec.~II is 
\textit{locally} isomorphic to the direct product group $\GSM$. 
There exist two alternative constructions which give locally isomorphic 
and apparently different groups. 

\begin{enumerate}
\item  A factor group of $\GSM$ by a discrete subgroup 
$\itGamma$ \,: \, $\tilde{G}_{\textrm{SM} }:= G_{SM}/\itGamma$ 

There are several discrete normal subgroups of $\GSM$. 
Among them we choose the cyclic group  $\mathbb{Z}_6$ generated by 
$g_{\omega}:= (\omega_3, \omega_2, \omega_6) \in G_{\textrm{SM}}\;\left(\omega_n = e ^{2\pi i/n} \right)$. 
The factor group 
\begin{equation}
	\GtSM:= G_{\textrm{SM}} /\mathbb{Z}_6
\end{equation}
yields the hypercharge selection rule. 
Indeed, a representation $r$ of $\GSM$ which corresponds to 
a representation $\tilde{r}$ of  $\GpSM$ should satisfy  
\begin{equation}
	r \left( g\, {g_{\omega}}^k\right) = r \left( g \right)\quad (k = 1, .., 5). 
	\label{eq: inv_repr_1}
\end{equation}

\vspace{1mm}

\centerline{
	\xymatrix{
	G_{\textrm{SM}}  \ar[r]^r \ar[d]_{\pi}& GL(V)  \\
	\GtSM \ar[ur]_{\tilde{r}}& }
}

In order to see the selection rule, 
let us consider  the one-parameter subgroup of $\GSM$, 
$g(\theta)=\left(\gamma_3(\theta),\gamma_2(\theta),e^{i\theta/6}\right)$ 
satisfying 
$g(\theta + 2\pi) =  g(\theta)\,g_{\omega}$. 
Equation~\eqref{eq: inv_repr_1} means 
\begin{equation}
	 r\left( g(\theta + 2\pi)\right) = r\left( g(\theta)\right). 
	\label{eq: inv_repr_2}
\end{equation}
Since the representation $r$ of $\GSM$ is of the form 
$r(g) = R_3(g_3)\otimes  R_2(g_2) \otimes  e^{iy\theta}$, 
we find 
\begin{equation}
	\begin{split}
	  r\left( g(\theta + 2\pi)\right) & =  R_3\left(g_3(\theta + 2\pi)\right) 
        \otimes  R_2\left(g_2(\theta + 2\pi)\right) 
           \otimes  e^{iy(\theta + 2\pi)}\\
 	    &=  e^{-2\pi i \left( \frac{r_2}{2}- \frac{r_3}{3}\right)} 
                           e^{2\pi iy} r\left( g(\theta)\right) .
	\end{split}
\end{equation}
From this equation and Eq.~\eqref{eq: inv_repr_2} we get 
\begin{equation*}
	y \equiv \frac{r_2}{2} - \frac{r_3}{3} \mod {\mathbb{Z}},
\end{equation*}
which is the same as Eq.~\eqref{eq:hypercharge}.

\item  The $SU(5)$-subgroup of the form 
\[
	G^{(5)} := \left\{
      \left(\begin{array}{cc} U_3 & 0 \\
                                     0 & U_2
            \end{array}\right)
\;\Bigr|\;
U_3 \in U(3), U_2 \in U(2),\, \det U_3 \det U_2 = 1 
    \right\}.
\]
This group can be brought about through the breakdown of $SU(5)$ with 
a non-zero vacuum expectation value of the Higgs boson belonging to 
the adjoint representation. 
\end{enumerate}

The seemingly different groups $\GpSM, \GtSM$ and $G^{(5)} $ are, in fact, 
isomorphic to each other as shown in the following commutative 
diagram \cite{Helgason}. 
\vspace{2mm}

\centerline{
	\xymatrix{
	 & G_{\textrm{SM}}  \ar[dl]_{\varphi_1}\ar[d]^{\pi} \ar[dr]^{\varphi_2}  \\
	 \GpSM \ar@/_1.5pc/ [rr]_{\phi_3} ^{\sim} & 
        \GtSM \ar[l]^{\phi_1}_{\sim}\ar[r]_{\phi_2}^{\sim} & 
        G^{(5)} & \!\!\!\!\!\!\!\! \subset SU(5)  \textrm{ (as a subgroup)}}
}

\vspace{1mm}

The isomorphism  $\GpSM \cong \GtSM$ is confirmed by the surjective homomorphism 
\begin{equation}
\begin{split}
   \varphi_1 &: \quad G_{\textrm{SM}} \quad \longrightarrow \quad \GpSM \\
     & \left(g_3, g_2 ,e^{i\theta} \right) \mapsto 
       \left(g_3 \gamma_3( -6\theta),\, g_2 \gamma_2(-6\theta),\, e^{6i\theta} \right),
\end{split}
\end{equation}
whose kernel is
\begin{equation}
   \Ker \,\varphi_1 = \left\{ {g_{\omega}}^k \right\}_{k = 0, \ldots ,5} 
         = \mathbb{Z}_6, \qquad g_{\omega}:= 
       ({\omega_3}\mathbf{1}_3, \omega_2 \mathbf{1}_2, \omega_6). 
\end{equation}
The isomorphism $G^{(5)} \cong \GtSM$ is verified by the surjective homomorphism 
\begin{equation}
	\begin{split}
	  \varphi_2 &: \quad G_{\textrm{SM}} \quad \longrightarrow \quad G^{(5)} \\
       & \left(g_3, g_2, e^{i\theta} \right) \mapsto 
           \left(g_3  e^{-2 i\theta},\, g_2  e^{3i\theta} \right) ,
	\end{split}
\end{equation}
whose kernel is
\begin{equation}
   \Ker \,\varphi_2 = \left\{ {g_{\omega}}^k \right\}_{k = 0, \ldots ,5} 
       = \mathbb{Z}_6, \qquad g_{\omega}:= 
         ({\omega_3}\mathbf{1}_3, \omega_2 \mathbf{1}_2, \omega_6). 
\end{equation}
The isomorphism  $\phi_3 :  \GpSM \rightarrow G^{(5)}$ is given by 
\begin{equation}
   \phi_3 =\phi_2 \circ {\phi_1}^{-1} : \, 
       \left(g_3, g_2, e^{i\theta} \right) \mapsto  
         \left(g_3 \gamma_3(\theta)e^{-\frac{i}{3}\theta}, \,
              g_2 \gamma_2(\theta)e^{\frac{i}{2}\theta}\right). 
\end{equation}

As discussed in Sec.~IV, in the theory with $\GpSM$ the gravitational anomaly cancellation 
is automatically guaranteed, while in the Standard Model with $\GSM$ it is not. 
This could be understood from the above-mentioned isomorphism between $\GpSM$ and the $SU(5)$-subgroup $G^{(5)}$.
The cancellation of the gravitational anomaly in $G^{(5)}$ is due to 
the traceless feature of the generators of a simple group $SU(5)$.

\section{EXOTIC MATTER FIELDS IN THE CONTEXT OF $\GpSM$}

In this section we consider possible matter fields beyond quarks an leptons 
in the Standard Model. 
In the context of 
$\GpSM = \left[SU(3)_{\rm c} \times SU(2)_{\rm L}\right]\rtimes U(1)_{\rm Y}$ 
gauge model, 
the $y$-charges of the matter fields are given by Eq.~\eqref{eq:hypercharge}. 
At the same time, their $y$-charges should be consistent with the anomaly-free 
conditions. 
Since it is difficult  to find its general solution, 
we restrict the present study to a simple example.

Here we take up an example of $\GpSM$ gauge model with exotic matter 
fields listed as 
\[
	q'_{\rm L} : (\bm{N},\ {\bf 2}), \qquad 
		u'_{\rm R},\ d'_{\rm R} : (\bm{N},\ {\bf 1}), \qquad 
	l'_{\rm L} : ({\bf 1},\ {\bf 2}), \qquad 
		e'_{\rm R} : ({\bf 1},\ {\bf 1}).
\]
It is seen below that this set is intriguing. 
In fact, if these matter fields cancel the gauge anomaly, then they automatically cancel both the $SU(2)$ global anomaly and the gravitational anomaly. In addition, we find that $|y(e_R)|$ has a rather large value $|y(e_R)| \geq 5$ except for the case $N=3$.

The $N$-plet of $SU(3)_{\rm c}$ associated with the highest weight $\lambda$ 
is characterized by the Dynkin label pair 
$[\lambda_1, \ \lambda_2]_{\rm D}$ of $\lambda$. 
Concretely, $N$ is expressed as 
\begin{equation}
	N = \frac{1}{2} (\lambda_1 + 1) (\lambda_2 + 1) 
		 (\lambda_1 + \lambda_2 + 2). 
\end{equation}
In the Young tableau the lengths of the first and the second rows are 
$\lambda_1 + \lambda_2$ and $\lambda_2$, respectively. 
The $y$-charges of the matter fields are given by 
\begin{equation}
	y(q'_{\rm L}) \equiv \frac{1}{2} - \frac{r_3}{3}, \quad 
	y(u'_{\rm R}) \equiv y(d'_{\rm R}) \equiv - \frac{r_3}{3}, \quad 
	y(l'_{\rm L}) \equiv \frac{1}{2}, \quad 
	y(e'_{\rm R}) \equiv 0 \quad \mod{\mathbb{Z}},
\end{equation}
where $r_3 = \lambda_1 + 2 \lambda_2$. 
In the present case the mixed anomaly conditions become 
\begin{equation}
	\begin{split}
 	& 2 y(q'_{\rm L}) - y(u'_{\rm R}) - y(d'_{\rm R}) = 0,  \\
 	&                N y(q'_{\rm L}) + y(l'_{\rm L}) = 0.  
	\label{eq:mixedN} 
	\end{split}
\end{equation}
In order that Eq.~\eqref{eq:mixedN} holds, $N$ should be odd and 
$r_3 N \equiv 0 \mod{3}$. 
The result that $N$ is an odd number is in line with the SU(2) global anomaly 
condition pointed out in \cite{Witten}. 
The cubic anomaly condition is 
\begin{equation}                
	2N y(q'_{\rm L})^3 - N y(u'_{\rm R})^3 - N y(d'_{\rm R})^3 
		+ 2 y(l'_{\rm L})^3 - y(e'_{\rm R})^3 = 0. 
\end{equation}
With the notation $y(u'_{\rm R}) = x$ and $y(d'_{\rm R}) = z$, 
this condition is translated as 
\[
	\frac{N}{4}(N^2-1) (x + z)^3 + N x^3 + N z^3 + y(e'_{\rm R})^3 = 0. 
\]
In the same manner as the study in Sec.~IV, through the transformation 
\begin{align}
	& x = \frac{1}{2N} ( -(N-1)X + (N+1)Z ), \nonumber \\ 
	& z = \frac{1}{2N} ( (N+1)X - (N-1)Z ), \nonumber 
\end{align}
the cubic condition becomes
\begin{equation}
	X^3 + Z^3 + y(e'_R)^3 = 0. \label{eq:Fermatprime}
\end{equation}
Denoting the $y$-charge of $q'_{\rm L}$ by 
\begin{equation}
	y(q'_{\rm L}) = \frac{1}{2} - \frac{r_3}{3} + n', \qquad n' \in \mathbb{Z}, 
\end{equation}
we get 
\[
	x = -\frac{r_3}{3} + n' + s', \qquad 
	z = -\frac{r_3}{3} + n' - s' + 1, \qquad s' \in \mathbb{Z}. 
\]
It follows that 
\[
	X = N y(q'_{\rm L}) - s' + \frac{1}{2}, \qquad 
	Z = N y(q'_{\rm L}) + s' - \frac{1}{2}.
\]
It is easy to see that if $y(e'_{\rm R}) = 0$, there is no solution of 
Eq.~\eqref{eq:Fermatprime}. 
In the case of $X = 0$ we obtain 
\begin{align}
	& y(u'_{\rm R}) = (N+1) y(q'_{\rm L}), \qquad 
		y(d'_{\rm R}) = -(N-1) y(q'_{\rm L}), \nonumber \\
	& y(l'_{\rm L}) = -N y(q'_{\rm L}), \qquad \qquad 
		y(e'_{\rm R}) = -2N y(q'_{\rm L}). 
\end{align}
In the case of $Z = 0$ the result is the same as the case of $X = 0$. 
This solution automatically satisfies the condition for 
the gravitational anomaly cancellation. 
On the other hand, in the Standard Model with $\GSM$ we find two types of solutions, i.e., $XZ = 0$ or $y(e'_R) = 0$, which satisfy the gravitational anomaly condition.  In this case we can not rule out one of the solutions: $y(e'_R) = 0$.
As the $N$-plet of $SU(3)_{\rm c}$ we can take $N = 3, \ 15, \ 21, \ 27, \cdots$. 
However, any solutions other than $N = 3$ and $n' = 0$ yield 
$|Q_{\mathrm{em}}(e'_R)| = |y(e'_R)| \geq 5$. 
It is worth pointing out that the ratios of the $y$-charges of matter fields 
relative to $y(q'_{\rm L})$ vary depending on $N$. 
Consequently, through the experimental study of the strength of $U(1)_{\rm Y}$ 
couplings with matter fields we are able to search the exotic fields 
for their evidence.

\section{SUMMARY}

The quantization of the $y$-charge does not occur in the Standard Model 
gauge group $\GSM$. 
In the direct product gauge group $\GSM$ the general solution to the anomaly-free 
conditions on the $y$-charges contains two real parameters. 
Therefore, within the context of the Standard Model with $\GSM$ 
there could be infinitely many non-standard $y$-charge assignments. 
In  $\GSM$, in general, the gravitational anomaly is not automatically canceled 
and its cancellation gives an additional 
condition on the $y$-charges. 

On the other hand, in the semidirect product gauge group 
$\GpSM = [SU(3)_{\rm c} \times SU(2)_{\rm L}] \rtimes U(1)_{\rm Y}$ 
the $y$-charges are quantized. 
This is because the semidirect product gauge group $\GpSM$ entails 
the selection rule beyond $\GSM$. 
The selection rule implies that the $y$-charge of each matter field has 
an intrinsic connection with the irreducible representation of 
$SU(3)_{\rm c} \times SU(2)_{\rm L}$ and also is settled to 
be $n/6 \mod {\mathbb{Z}}\;(n = 0,1,3,4) $. 
The gauge-anomaly-free conditions by itself strongly constrain the $y$-charges and 
their ratios are uniquely determined. 
Furthermore, in  $\GpSM$ the gravitational anomaly cancellation is automatically guaranteed. 
This is linked to the fact that $\GpSM$ is isomorphic to an $SU(5)$-subgroup. 
In addition, under the constraint 
$| Q_{\rm em}(e_{\rm R}) | = | y(e_{\rm R}) | < 5$ 
we attain to the standard charge assignment of quarks and leptons.

Finally, a few remarks may be in order. 
Although the gauge transformation rules of the gauge fields in $\GpSM$ is 
slightly different from those in $\GSM$, the Lagrangian in $\GpSM$ is the 
same as in $\GSM$. 
We pointed out that three gauge groups constructed in the distinct manners, 
i.e., $\GpSM$, $\GtSM$ and $G^{(5)}$, are isomorphic to each other. 
For exotic matter fields, if they exist, their $y$-charges are predicted. 
Through the experimental study of the strength of $U(1)_{\rm Y}$ couplings 
with matter fields we are able to search the exotic fields 
for their evidence. 

\vspace{10mm}


\section*{ACKNOWLEDGEMENTS}
The authors are grateful to Masato Ito for a valuable comment on 
the gravitational anomalies. 

\vspace{5mm}



\end{document}